\documentclass[runningheads]{llncs}
\usepackage[T1]{fontenc}
\usepackage{graphicx}
\usepackage[hidelinks]{hyperref}
\usepackage{color}

\usepackage[square, numbers]{natbib}
\begin{document}
\title{Nano-ESG: Extracting Corporate Sustainability Information from News Articles} 
\titlerunning{Nano-ESG: Extracting ESG Information from News Data}
%
\author{Fabian Billert \and
Stefan Conrad}
\institute{Heinrich-Heine University of D\"usseldorf, D\"usseldorf, Germany\\
\email{\texttt{\{fabian.billert, stefan.conrad\}@hhu.de}}}
\maketitle              
\begin{abstract}
Determining the sustainability impact of companies is a highly complex subject which has garnered more and more attention over the past few years. Today, investors largely rely on sustainability-ratings from established rating-providers in order to analyze how responsibly a company acts. However, those ratings have recently been criticized for being hard to understand and nearly impossible to reproduce.

An independent way to find out about the sustainability practices of companies lies in the rich landscape of news article data. In this paper, we explore a different approach to identify key opportunities and challenges of companies in the sustainability domain. We present a novel dataset of more than $840,000$ news articles which were gathered for major German companies between January 2023 and September 2024. By applying a mixture of Natural Language Processing techniques, we first identify relevant articles, before summarizing them and extracting their sustainability-related sentiment and aspect using Large Language Models (LLMs). Furthermore, we conduct an evaluation of the obtained data and determine that the LLM-produced answers are accurate. We release both datasets at \url{https://github.com/Bailefan/Nano-ESG}. 

\end{abstract}
\section{Introduction}
With the growing fear of climate change related issues in the recent past, sustainability has become an important topic not only for the general public, but also in the corporate world \cite{zumente.2021}, where it is coined as ``ESG'', standing for Environmental, Social and Governance topics. Those involved in the corporate world, both on the investor and on the company side, desire clarity about how sustainable a company is \cite{islam.2020, Zumente.2021b}.

Third party rating agencies offer products which aim to represent the sustainability of a company in a single value, an ``ESG-score''. While convenient for investors, recent criticism has pointed out several problems with this approach. For one, different rating agencies can arrive at a variety of scores for the same company, which confuses users and raises concerns about the derived scores \cite{Berg.2019}. In addition, the various complex steps involved in determining a single value for a company make it hard for the end user to truly understand why a company has a certain ESG-rating \cite{Clément.2023}.

With the recent advancements in the field of Natural Language Processing (NLP) and the rise of Large Language Models (LLMs), analyzing vast amounts of data has become more feasible \cite{Ouyang.2022,OpenAI.2023,Touvron.2023, Touvron.2023b}. One textual data source which offers important insights for corporate sustainability practices are news articles. They provide a real-time, unfiltered source of information that reflects public perception and current events related to corporate sustainability practices. Unlike formal reports or company disclosures, which may be delayed or subject to selective reporting, news coverage can capture immediate reactions to a company’s actions, controversies, or initiatives \cite{salmones.2021, Dorfleitner.2024}. Leveraging NLP techniques in order to extract the ESG-information available in this public data source can enable different stakeholders to track corporate sustainability more transparently, without relying solely on ratings that may obscure the complexity of ESG performance \cite{Oliveira.2024}. Furthermore, it is possible to trace a sentiment back to a concrete source and thus evaluate each ESG event on its own if desired. Finally, this approach also enables continuous monitoring, capturing shifts in perception that would otherwise go unnoticed between formal ESG score updates.

To further research in this direction, we analyze a large number of online news articles in order to extract information relevant to a company's ESG practices in this work. By leveraging multiple modern Machine Learning (ML) models to filter out as much noise as possible, we can process a small amount of news articles with a powerful, but expensive LLM to create a rich dataset we call \texttt{Nano-ESG} - named for the level of detail each company can be investigated with. This dataset currently contains companies in the German DAX 40 financial index, as well as a few companies that recently dropped out of it. After determining if a news article is relevant for a given company in the context of sustainability, we summarize it and determine the sentiment as well as which ESG-aspect (meaning Environmental, Social or Governance) it belongs to. To maximize the quality of the final responses, we use GPT-4o\footnote{\url{https://openai.com/index/hello-gpt-4o/}}.

\texttt{Nano-ESG} contains news articles from both German and English media sources from January 2023 until September 2024. By further leveraging LLMs, the dataset contains summaries in both German and English. While we cannot release the original news articles themselves for copyright reasons, we release the URLs of each news article along with the concerned company, the discussed summaries, the determined ESG-sentiment and the ESG-aspect. As the publication date of each article is also included, this is, as far as we know, the first open source dataset which contains a time-series of ESG-sentiments and ESG-aspects of companies, which can enable practitioners to analyze the interplay of ESG-events with external factors.

In order to be able to judge the reliability of our dataset and the quality of the final responses, we conduct a rigorous evaluation of a subset of the generated data in \autoref{sec:eval}. We further carry out a detailed analysis of the dataset in \autoref{sec:analysis} and present a use case for identifying ESG-topics. Finally, we analyze unsatisfactory responses, examine why they were generated and present our findings in \autoref{sec:error_analysis}.

\section{Related Work}

\textbf{Large Language Models for Data Annotation} In the last few years the capabilities of both closed and open source LLMs have skyrocketed, with novel LLMs being capable of complex reasoning tasks \cite{Touvron.2023, Touvron.2023b, OpenAI.2023, llama3.1.2024}. As a result, LLMs have become popular as tools to process textual data and extract desired properties in an automatic way \cite{Tan.2024}. \cite{Ding.2023} study different approaches in which a GPT-3-based model, \texttt{text-davinci-003}, is used to annotate various datasets. The authors come to the conclusion that GPT-3, although still lacking behind human annotators, shows potential in performance while being significantly cheaper. \cite{Nasution.2024} investigates the performance of the improved GPT-4 in annotating low-resource languages and finds impressing results of the LLM, breaking even or outperforming human annotators in many cases. \cite{He.2024} compares GPT-4 with an elaborate Amazon Mechanical Turk (MTurk) setup to annotate a subset of the CODA-19 dataset \cite{Huang.2020rla} and finds that GPT-4 achieves a higher performance. \cite{Törnberg.2023} comes to a similar conclusion when analyzing annotations of political twitter messages from \cite{Vliet.2020} between GPT-4, MTurk workers and Experts.

\textbf{Datasets for ESG-Research} Datasets in the ESG domain are scarce, but have been gaining interest recently. The FinNLP series of workshops\footnote{\url{https://aclanthology.org/venues/finnlp/}} has published multiple ESG-centric and organized shared tasks for researchers. They have focused on identifying ESG-aspects \cite{chen2023ESG}, ESG-sentiments \cite{Chen.2023b} and ESG impact duration \cite{Chen.2024}.

Another noteworthy dataset was created by \cite{Fischbach.2023}. Here, the authors gathered more than $450000$ headlines from the twitter account of \textit{The Guardian}\footnote{\url{https://www.theguardian.com/}}, checked the tags for ESG-relation and mapped relevant headlines to ESG-aspects manually. They then trained models to classify the Relevancy, ESG-aspect and ESG-sentiment of the headlines. Notably, this work does not include timestamps, making it difficult to place ESG-events in context with other events. Additionally, it focuses on the article headlines as opposed to the full article content. Finally, in a different approach, \cite{Pavlova.2024} creates a corpus of nearly $4000$ news articles about $10$ companies in the FTSE financial index with ESG-relevance as well as ESG-aspects labels. This work uses the whole article content for their classification approaches but also does not contain timestamps.

\section{Data Processing Pipeline}

\begin{figure}[h]
    \centering
    \includegraphics[width=1\textwidth]{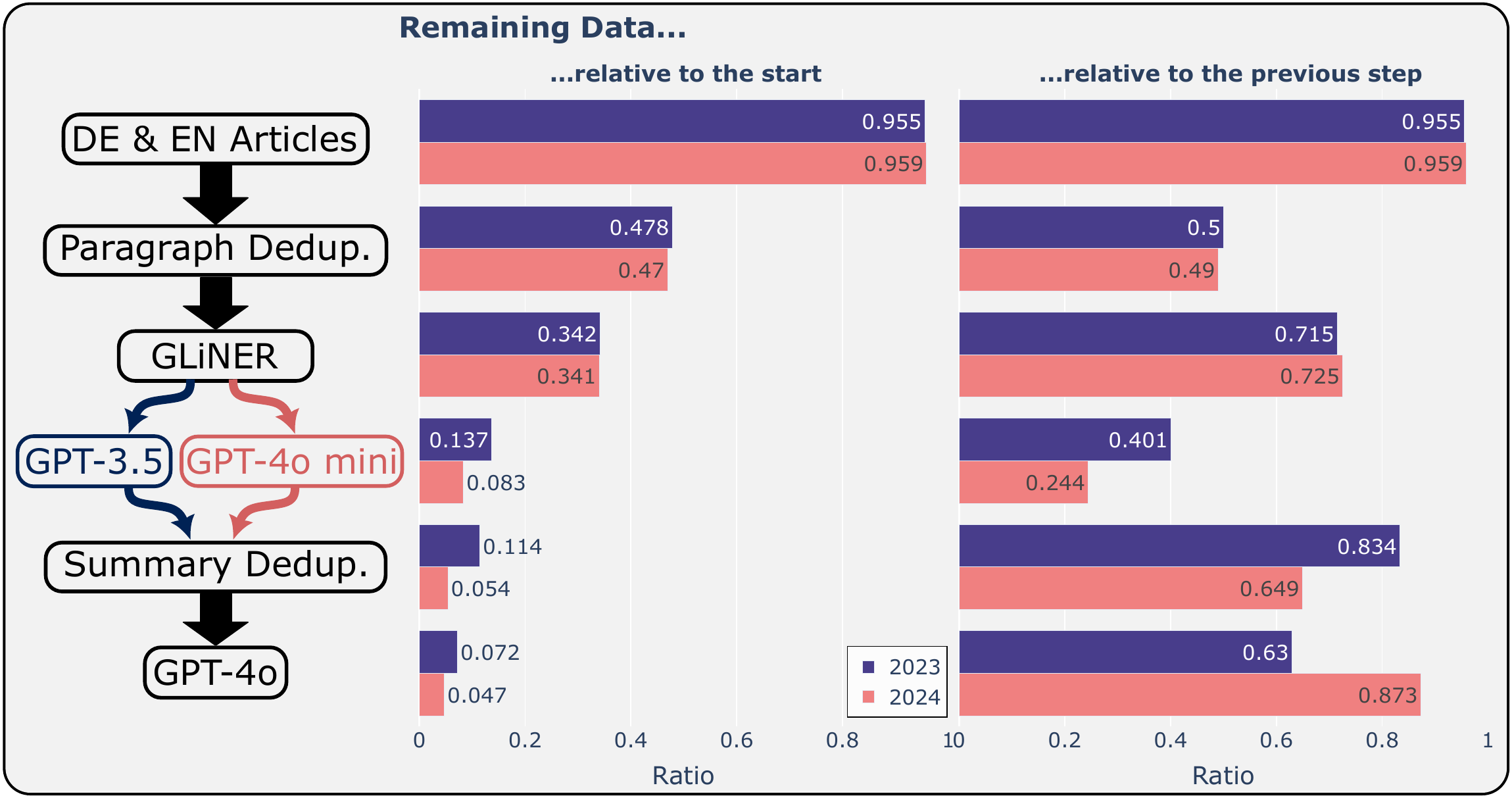}
    \caption{Processing-pipeline (left) used for the creation of the dataset and how much each step reduces the amount of data relative to the starting point (middle) and relative to the previous step (right).}
    \label{fig:data_pipeline}
\end{figure}

In this section, we describe the pipeline used to obtain the final information contained in the \texttt{Nano-ESG} dataset. To reduce the cost of processing the articles with the more expensive LLM in the last step as much as possible, we apply different techniques to filter out articles unlikely to be relevant before then. A schema of our workflow along with how much of the initial data is filtered out (compared to the total at the start as well as each previous step) is displayed in \autoref{fig:data_pipeline}. As the data processing pipeline for $2024$ differs slightly from the one for $2023$, we display them separated in two colors in the graph.
\vspace{-2mm}
\subsection{Raw Data}

The original data is obtained by crawling popular English and German online media websites. Social media sources such as twitter and Facebook are not considered. As this dataset is currently constrained to German companies, we defined a list of keywords\footnote{See \texttt{company\_info.py} in our Github repository for the full list} for each company in the German DAX 40 financial index and search for articles which contain any of these keywords. After this, we are left with around $850,000$ articles for the considered timeframe from January 2023 until the $16$th of September 2024. Some of the crawled articles were in other languages - we decided to proceed with only English and German articles for simplicity and lower costs.
\vspace{-2mm}
\subsection{Paragraph Deduplication}
\label{subsec:dedupe1}
During the initial data exploration we noticed that online sources often publish similar articles for a specific event. In order to condense the displayed information, we employ an embedding model to find similar articles in a set timeframe. We use a bilingual embedding model which was trained on English and German data \cite{Mohr.2024}, \texttt{jina-embeddings-v2-base-de}\footnote{\url{https://huggingface.co/jinaai/jina-embeddings-v2-base-de}} to embed the paragraphs of the articles which contain any one of the company-specific keywords. We then group together articles published within a week of each other for which we calculate a cosine-similarity of $0.8$ or more. We decided on this value after experimenting with different cutoffs for various groups of articles. Once we find articles similar to one another, we keep the first article in the dataset and discard the following articles. This step allows us to reduce the amount of articles by around $50\%$.
\vspace{-2mm}
\subsection{Entity-based Filtering}

Some companies have associated keywords which can also refer to unrelated information. To deal with these cases, we use a multilingual zero-shot named entity recognition model called GLiNER, more specifically \texttt{gliner\_multi-v2.1} \cite{Zaratiana.2023}. To filter out articles which do not refer to any of the companies we are interested in (for example where ``Allianz'' refers to a pact between entities rather than the German company), we apply this model to the sentences in which we detected a keyword. We discard the article if none of the keywords refer to an organization or company.

\vspace{-2mm}
\subsection{Filter Language Model}
\label{sec:filter_step}
As depicted in \autoref{fig:data_pipeline} in the middle plot, the previously described techniques can reduce the amount of data by around two thirds. However, in order to reduce costs further, we decided to first use a relatively cheap language model to filter out irrelevant articles with regards to either the target company or ESG thematics. We chose to use OpenAI's \texttt{gpt-3.5-turbo-0125} for its ease of use and competitive pricing. While we were working on processing the data, a more performant and cheaper model, \texttt{gpt-4o-mini-2024-07-18} was released, so we switched to this model instead. The result is that for this step, data from 2023 was processed with GPT-3.5-turbo, and data from 2024 was processed with GPT-4o-mini.

The prompt we use is included in the supplementary material. In an attempt to force the model to reflect about its answer similar to a chain-of-thought approach \cite{Wei.2022}, we ask it to first generate a first answer for the relevancy of the article. We then prompt it to explain its answer as well as generate a summary for the article, which we will use in a later step. We noticed that the model often inferred a relevance for ESG issues from loosely related topics (e.g. reasoning that a change of the stock price can have effects on the governance of the company), so we also ask the model to determine if ESG issues are being directly addressed in the article. After this, we let the model determine the relevancy again and refer it to its own answers for the explanation, summary and the direct ESG connection. Finally, we continue with all articles which directly address ESG issues and were judged to be \texttt{relevant}. This step manages to reduce the amount of data the most - by $60\%$ when using GPT-3.5-turbo in $2023$ and by more than $75\%$ with GPT-4o-mini in $2024$. Although the underlying data is different for the two models, it seems that the latter LLM is more adept at determining relevant articles.
\vspace{-2mm}
\subsection{Summary Deduplication}

At this point, we still have some articles in our dataset which are quite similar to one another. To further reduce the cost of the final determination of each article's properties, we use the summary generated in the previous step in order to remove similar articles with the same approach as in \autoref{subsec:dedupe1}. Again, we can observe a significant difference as to how  much data is filtered out by this step - for $2024$ nearly $20\%$ more articles are removed here compared to $2023$. A reason for this could be the worse relevancy-detection in the previous step by the older model, allowing more irrelevant articles of varying topics to still be in our dataset at this point.
\vspace{-2mm}
\subsection{Final Determination Model}
\label{sec:final_determination}

In the final step we decide to use a more powerful language model to create high quality data which can potentially be used in order to train small, but specialized models in the future. We chose the model based on the chatbot arena \cite{Chiang.2024}, which evaluates Large Language Models based on human preference. When we were working on this step, the highest performing model there was the GPT-4o model from OpenAI, more specifically \texttt{gpt-4o-2024-05-13} - all final responses are generated with this model. At the time of writing this, a few other models perform better, but the version of GPT-4o we used still places in the top ten.

In this step we again ask the model to summarize the article regarding ESG-topics about the target company as well as determine if the article is relevant or not. If it is, we also ask for the ESG-sentiment, meaning the sentiment of the ESG-information for the target company, and the ESG-aspect. Again, the full prompt used in this step is included in the supplementary material. Although we instruct the model to reply in German, we decided to formulate the prompt in English since this results in less tokens and OpenAI uses a pay-per-token pricing model. To the best of our knowledge, this does not degrade the performance. Finally, in order to make sure the generated summaries do not violate German copyright law, we shorten and rephrase them.

This step is the most expensive step of our pipeline, so our goal was to have as few irrelevant articles remaining as possible. As shown on the right in \autoref{fig:data_pipeline}, the newer GPT-4o-mini left significantly less irrelevant articles in the dataset compared to the older model - only around $13\%$ of the articles processed by GPT-4o were deemed to be irrelevant.

\begin{figure}[h]
    \centering
    \includegraphics[width=1\textwidth]{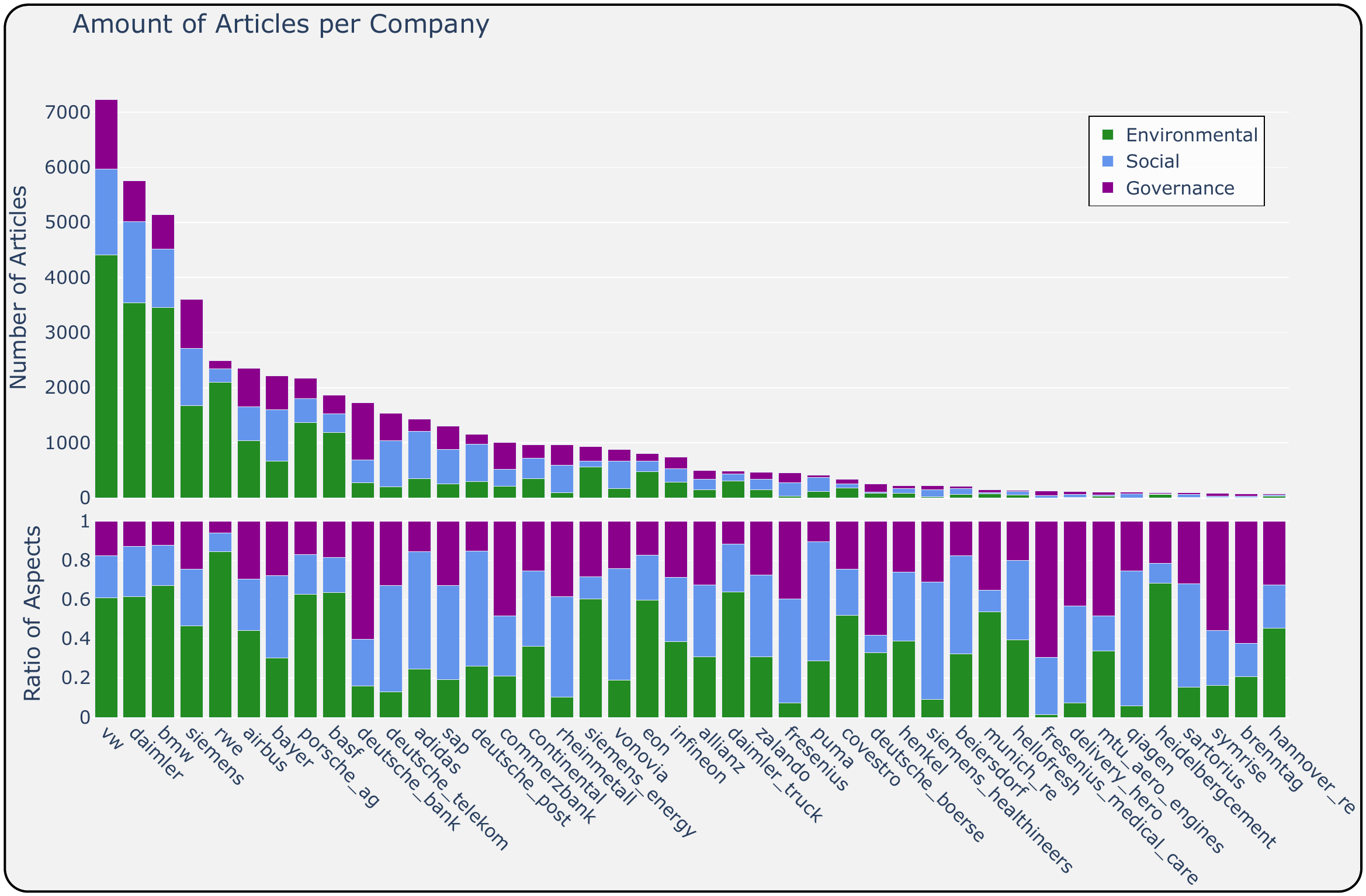}
    \caption{Number of articles with different ESG-aspects per company. Top: total number. Bottom: Ratio of each aspect per company.}
    \label{fig:dataset}
\end{figure}

\section{Nano-ESG}

\begin{table*}[h]
\centering
\caption{Top row: The total amount of articles for each sentiment and ESG-aspect. Bottom row: The ratio of articles per sentiment and ESG-aspect averaged for all companies.}
\begin{tabular}{ |p{4cm}|p{1.2cm}p{1.2cm}p{1.2cm}||p{1.2cm}p{1.2cm}p{1.2cm}|  }
\hline
& Negative & Neutral & Positive & E & S & G \\
\hline
Total & 17668 & 6115 & 27304 & 24546 & 15086 & 11455  \\
Avg. Perc. per Company & 0.3274 & 0.1290 & 0.5436 & 0.3574 & 0.3430 & 0.2995  \\
\hline
\end{tabular}
\label{tab:dataset}
\end{table*}
\vspace{-2mm}

After analyzing roughly $850,000$ articles, the remaining dataset consists of $51,087$ relevant articles with ESG-information. Each article has a timestamp of publication, a URL as well as the fields mentioned in \autoref{sec:final_determination}. In addition, we also asked the model to score the relevancy of the article and determine a list of keywords it deems important for each article. We will investigate these fields in \autoref{sec:analysis}. Finally, we also provide an English version of the summary which was created by translating with \texttt{gpt-4o-mini-2024-07-18}.

Each article is associated with one company. An article can appear multiple times in the final dataset, but the summary differs as it is always composed with regards to a single target company. \autoref{tab:dataset} shows the total number of articles categorized by sentiment and ESG-aspect, along with the distribution of these categories averaged over all companies. While the distribution of ESG-aspects per company is approximately similar, the amount of neutral articles is significantly lower than negative and positive articles. At the same time, as is shown in the top part of \autoref{fig:dataset}, it is worth noting that the total amount of articles between companies varies significantly. The fewest ESG-relevant articles we find for a company is $77$ articles for Hannover Re, while for the company with the most articles (Volkswagen) we find $7233$ relevant articles. In the bottom part of the figure we display the ratio of articles for each company which belong to each aspect. As can be suspected from the bottom row of \autoref{tab:dataset}, the companies with the most articles in general all have significantly more environmental articles than articles for other ESG-aspects.

\section{Data Evaluation}
\label{sec:eval}

In order to assess the quality of the generated results, we evaluated a subset of the data manually. We asked five sustainability experts to annotate the data and used Argilla as our annotation-interface\footnote{\url{https://docs.argilla.io/latest/}}. The results are evaluated in two steps: First, we check if the summaries were correctly created by GPT 4-o. In a second, separate task, we evaluate the sentiment and aspect determined by the LLM. The instructions for the the two annotation tasks are included in the supplementary material.

\vspace{-2mm}
\subsection{Evaluation Dataset}
To create a dataset to evaluate the summarization task, we randomly picked one article for each (company, sentiment) group, ending up with $123$ summaries. $38$ of these samples, roughly one per company, were assigned to all annotators.

For the more complex second task we decided to evaluate more samples. In order to pick diverse summaries, we applied k-means clustering (with $n=3$) to each (company, sentiment) subgroup using their embeddings and chose the nearest article from each cluster center. In total, $367$ samples were evaluated. Of these, $82$ articles (two per company, with approximately the same number of articles for all sentiments) were shown to all annotators.
\vspace{-2mm}
\subsection{Summary Evaluation}

We first analyze the samples all annotators evaluated. For $35$ of the $38$ samples, all annotators agreed on the correctness of the summary. In the other three cases, only one annotator disagreed. After consulting with the annotators, it became clear that the problem for the three cases lies in the article not containing any relevant ESG-information for the target company in the first place. Looking at all samples, in $95.9\%$ of the cases all annotators determined the summary to be \texttt{correct}. We conclude that the LLM generally is able to correctly summarize the article according to the instructions, but can struggle to assess its relevancy. We investigate this further in the second evaluation task.

\subsection{Classification Evaluation}

We discuss the different fields that were labeled separately:

\textbf{Relevancy:} Of the articles evaluated by all five sustainability experts, $68.7\%$ are deemed as relevant by all annotators. For $83.1\%$ of the samples, a majority of $3/5$ annotators decided the sample was relevant. Finally, $98.8\%$ of the samples were determined as relevant by at least one annotator.

\textbf{Sentiment:} Evaluating the sentiment raised some problems. Some annotators marked all sentiments appearing in parts of the summary while others simply marked the most relevant sentiment. We decided to simplify the sentiments according to the following rules: If an annotator selected \texttt{neutral} along with either \texttt{positive} or \texttt{negative}, we assign the non-\texttt{neutral} value. If an annotator selected both \texttt{positive} and \texttt{negative}, we set the sentiment to \texttt{neutral}, which we also do if they were \texttt{not sure}. We then measure Fleiss' kappa \cite{Fleiss.1971} for all samples all annotators deemed relevant to be $\kappa = 0.818$. From these findings it seems that the task of determining the ESG-sentiment for a summary is not straight-forward, but the annotators had a generally reasonable agreement as to the sentiment of a sample. The final sentiment was determined by taking the value picked by most annotators or following a similar approach as above if the values are tied. We then measure an accuracy of $79.9\%$ of the LLM annotations compared to the the user annotations.

\textbf{Aspects:} We further investigate the aspect evaluations. We again calculate Fleiss' kappa by taking all samples all annotators found relevant and arrive at a value of $\kappa = 0.427$. Annotators evidently found this task more difficult to judge than the sentiment. To determine the aspect the annotators most agree about for a sample, we first look at the aspect most selected among all annotators. If there is a draw, we look at the aspect the annotators deemed the most relevant individually and select the most selected aspect from there. From this, we calculate an accuracy of $78.5\%$ for the aspect determined by the LLM compared to the one determined by the annotators. Notably, the most disagreement appeared between samples the LLM determined as \texttt{social} while the annotators selected \texttt{governance}.

\section{Results}
\label{sec:analysis}
\vspace{-2mm}
\subsection{Development of the ESG-Aspects}
\begin{figure}[h]
    \centering
    \includegraphics[width=1\textwidth]{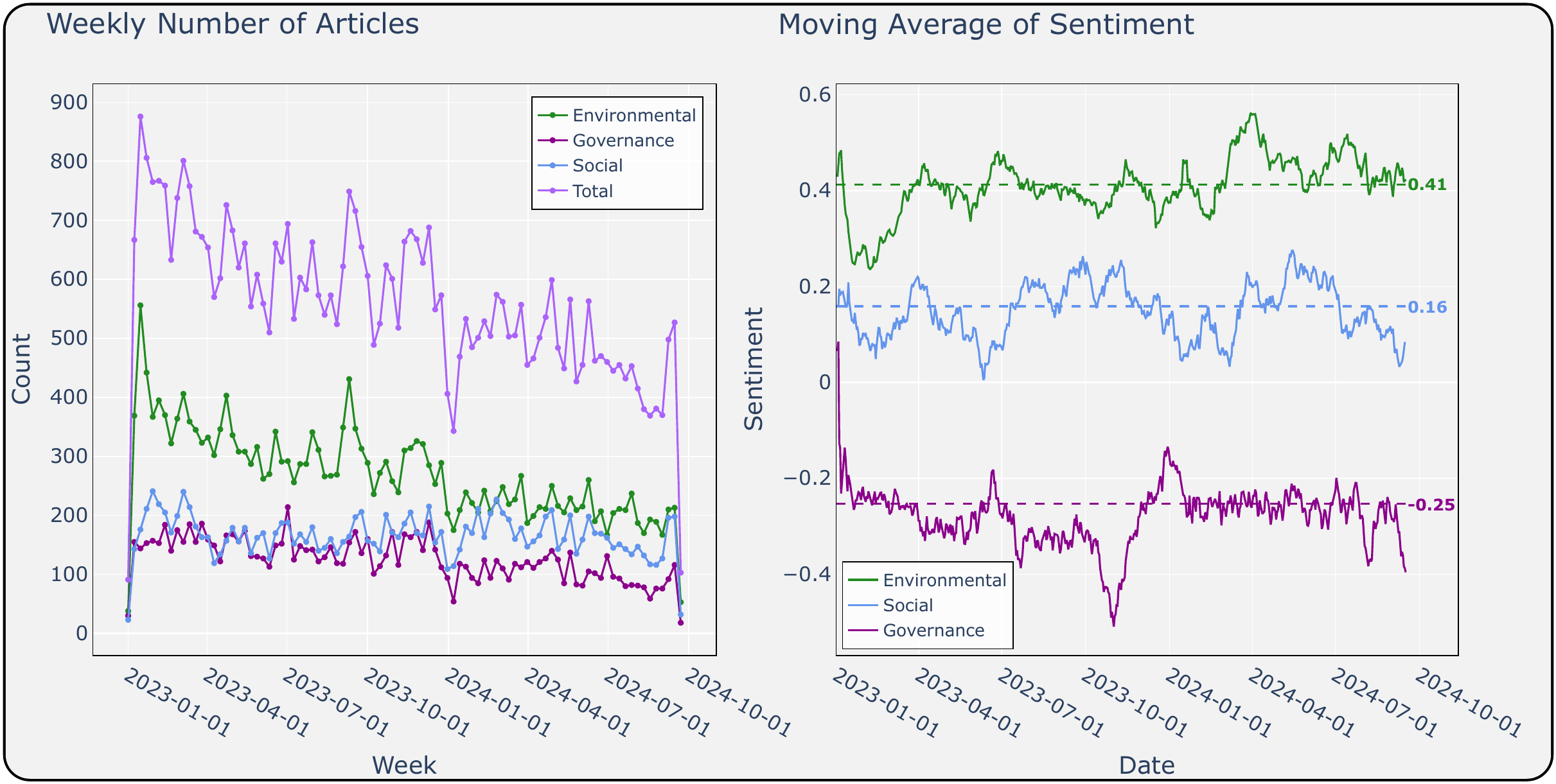}
    \caption{Left: Weekly number of articles in total and for each ESG-aspect. Right: 30-day moving average of the sentiment per ESG-aspect.}
    \label{fig:data_analysis}
\end{figure}
Over the $21$-month period investigated in this work, the total number of ESG-relevant news articles we detect for the observed companies generally decreases. As we can see in \autoref{fig:data_analysis} on the left, which shows the weekly number of news articles for each ESG-aspect, this decrease can mostly be attributed to the environmental and governance aspects while the number of social articles remains approximately stable.

In the right figure of \autoref{fig:data_analysis}, we first calculate the daily mean sentiment and then apply a $30$-day moving-average for each ESG-aspect. We find that the environmental aspect is generally the most positive with a mean sentiment of $0.41$. Articles with a social ESG-aspect are only slightly positive at a mean sentiment of $0.16$. Overall, governance is the only consistently negative aspect during our observed timeframe, with a mean sentiment of $-0.25$.

\vspace{-2mm}
\subsection{Can LLMs Judge Degrees of Relevance?}

\begin{table}[t]
\caption{Top: Amount of articles for each relevance score in the full dataset. Bottom (two rows): Articles deemed irrelevant by at least one annotator per Relevance Score as judged by the LLM (irrelevant/total articles and associated ratio).}
\centering
\begin{tabular}{ |p{2.7cm}|p{0.6cm}p{0.6cm}p{0.8cm}p{0.8cm}p{1.2cm}p{1.2cm}p{1cm}p{0.8cm}p{0.8cm}|  }
\hline
\textbf{Relevance Score} & \textbf{2} & \textbf{3} & \textbf{4} & \textbf{5} & \textbf{6}& \textbf{7} & \textbf{8} & \textbf{9} & \textbf{10} \\
\hline
\textbf{All Data} & 2 & 122 & 983 & 1380 & 10465 & 18219 & 17113 & 2780 & 23 \\
\hline
\textbf{Eval} - Total & - & - & 3/6 & 3/8 & 33/63 & 19/101 & 7/141 & 1/50 & -  \\
\textbf{Eval} - Ratio & - & - & 0.50 & 0.38 & 0.52 & 0.19 & 0.05 & 0.02 & -  \\
\hline
\end{tabular}
\label{tab:relevancy}
\end{table}

Since we also prompted the LLM to provide a relevance-score between $1$ and $10$ for each article, we investigate here whether the returned score is a meaningful measure of relevancy. The top row in \autoref{tab:relevancy} displays the total amount of articles for each relevance score as assigned by GPT-4o. Most of its responses are centered around the score of $7$ - the overall mean relevance score is $7.12$. The model only rarely responds with low scores - values of $1$ are never assigned, and there are just two cases of $2$. Similarly, the highest score of $10$ is only given to $23$ articles.

The bottom two rows of \autoref{tab:relevancy} show the annotation results in the evaluation set. In the middle row we show how often at least one annotator deems an article irrelevant out of the total number of articles present in the evaluation set. The corresponding ratio in the bottom row indicates that articles with a lower relevance scores are more likely to be determined as irrelevant. At the same time, higher relevance scores are rarely determined as irrelevant - for a score of $9$, there is only one out of $50$ annotated articles with this label. We conclude from these findings that the LLM is generally able to determine to which degree an article is relevant in the here investigated context of ESG topics.

\subsection{Detecting Relevant Topics}
\label{sec:topic}

\begin{figure}[!ht]
    \centering
    \includegraphics[width=0.9\textwidth]{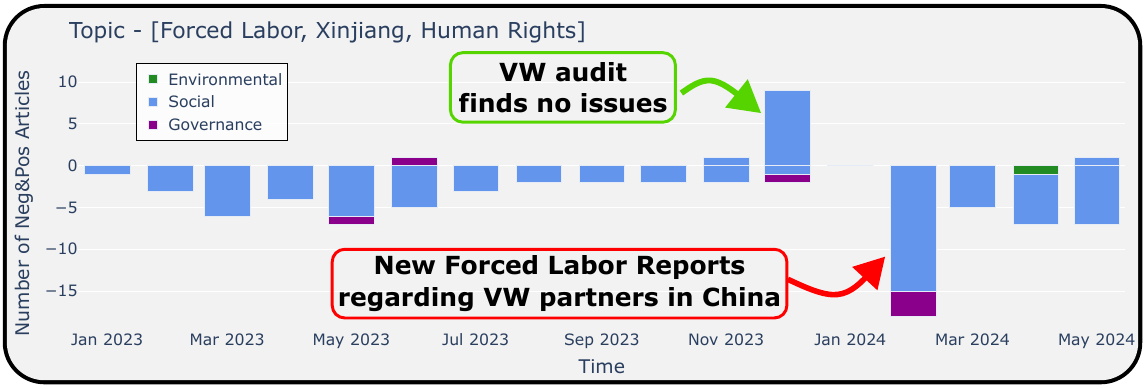}
    \caption{Amount of positive and negative articles regarding the ``Forced Labor'' topic of Volkswagen over time for the different detected aspects.}
    \label{fig:topic}
\end{figure}

We use \texttt{Bertopic}, a popular library for topic modeling, to identify relevant topics for each company \cite{Grootendorst.2022}. \texttt{Bertopic} identifies topics through the following steps: First, the dimensionality of the embeddings of the underlying documents is reduced before their clusters are detected. To determine good representations for the identified clusters, the default algorithm applies a bag-of-words approach for each cluster and then uses a modified TF-IDF technique to create a set of representing words per cluster.

In order to encourage the detection of sustainability-specific topics, we decided to augment the article summaries with the keywords the LLM determined in \autoref{sec:final_determination} while removing mentions of the companies which are sometimes included in them. We simply append the keywords to the end of the summaries and use the same embedding model as in \autoref{subsec:dedupe1} to embed the resulting text.

In the previous subsection, we identified the relevance score as a meaningful measure for the degree of relevancy of the articles. By calculating the mean relevance score of each topic cluster, we can thus determine relevant topics for the companies, as well as check if a topic is more likely to represent an opportunity or a risk through the mean sentiment. In addition, by looking at the amount of detected news articles for topics over time, it is possible to identify when critical events take place and how companies attempt to address them. Taking Volkswagen as an example, one of its most relevant topics we detected concerns reports of forced labor in China's Xinjiang province taking place in plants of the company. As shown in \autoref{fig:topic}, this is a generally negative topic for the company in 2023. In the end of 2023, an audit is conducted in its plants and finds no problems. However, shortly afterwards, in February 2024, new reports emerge, showing that the problem persists with Chinese partners of Volkswagen.

In this way, it is possible to explore different topics of the covered companies and even contextualize them with other real world events\footnote{A notebook in the supplementary material allows users to analyze the different topics of our covered companies further on their own.}.

\section{Limitations \& Error Analysis}
\label{sec:error_analysis}

During our evaluation, we identified several limitations in our current approach. One significant issue is the inability to differentiate conflicting sentiments within a single news article. Currently, we only capture an overarching sentiment, meaning that if multiple perspectives exist, we may only capture one, thus omitting valuable nuance.

Additionally, the LLM sometimes struggles to isolate ESG sentiment pertaining specifically to the target company, confusing it with the overall tone of the article. This is particularly problematic for reinsurance companies impacted by increased natural disasters. Such articles might be classified as (negative, environmental), which does not necessarily reflect the sentiment relevant to the reinsurance firm itself.

Another challenge is handling articles about multiple related companies. For instance, news about both \texttt{Siemens} and its subsidiary \texttt{Siemens Energy} may be marked as relevant to both companies, even when the article predominantly concerns one. Despite attempts to address this by including related companies in the prompt, this approach has had limited success.

\section{Conclusion \& Future Work}

We present \texttt{Nano-ESG}, a novel dataset containing corporate sustainability information extracted from news articles. By determining an ESG-sentiment and ESG-aspect for each relevant article, we are able to provide an alternative way to investigate corporate sustainability events - we present one approach for this through the detection of relevant topics in \autoref{sec:topic}. Our pipeline for information extraction is described in detail in this work and can be applied to other domains. 

We will continue to extend the work on this project by regularly adding more data to monitor the state of German corporate sustainability practices. In the future, we will aim to build an interactive application to allow users to conveniently access up-to-date information in this domain. In addition, we would like to extend the covered companies to more countries beyond Germany. To make this second goal feasible, it might be necessary to train more specialized models for the different data extraction steps, which should be possible due to the high-quality data gathered in this work. This would allow us to reduce costs further and thus increase the amount of data we can continuously monitor.

\begin{credits}
\subsubsection{\ackname} We want to thank GET Capital AG for the support in infrastructure they provided.
\end{credits}

\bibliographystyle{splncs04}  
\bibliography{Nano_dataset}

\end{document}